\RequirePackage{fix-cm}
\documentclass{svjour3}                     
\smartqed  

\usepackage{url}
\usepackage{graphicx}
\usepackage{subfigure}
\usepackage{booktabs}
\usepackage{natbib}
\usepackage{lineno}

\usepackage{latexsym}
\begin{document}

\title{Projected changes of rainfall seasonality and dry spells  in a high concentration pathway 21st century scenario  
}

\titlerunning{Future changes of rainfall seasonality }        

\author{Salvatore Pascale         \and
         Valerio Lucarini    \and
         Xue Feng    \and         
         Amilcare Porporato  \and
         Shabeh ul Hasson
}


\institute{\textbf{S. Pascale}, V. Lucarini, S. Hasson \at
             - KlimaCampus, Meteorologisches Institut, Grindelberg 5, Hamburg, 20144, Germany.\\
              \email{salvatore.pascale@uni-hamburg.de}           
           \and
           V. Lucarini \at
          -  Department of Mathematics and Statistics, University of Reading, Reading, UK \\ 
          -  Walker Institute for Climate System Research, University of Reading, Reading, UK
           \and
           X.Feng, A. Porporato \at
          - Department of Civil and Environmental Engineering,  Duke University, NC, USA   
}

\date{Received: date / Accepted: date}

\maketitle

\begin{abstract}
  In this  diagnostic  study we analyze  changes  of rainfall seasonality and dry spells   by the end of the  twenty-first century under the  most extreme IPCC5 emission scenario (RCP8.5)  as projected by twenty-four coupled climate models   participating to Coupled Model Intercomparison Project 5 (CMIP5). 
  We use  estimates of  the centroid of the monthly rainfall distribution  as an index of the rainfall timing and a threshold-independent,  information theory-based  quantity such as relative entropy  (RE) to quantify  the concentration of annual rainfall and the number of dry months and to build a monsoon dimensionless seasonality  index (DSI). The RE is projected to    increase,   with high inter-model agreement     over  Mediterranean-type regions -- southern Europe, northern Africa and southern Australia -- and areas of South and Central America,   implying  an  increase in the number of dry days up to one month by the end of the twenty-first century.   Positive  RE changes are also projected over the monsoon regions of southern  Africa and North America, South America. These trends are consistent with a shortening of the wet season   associated with  a more prolonged pre-monsoonal dry period. The  extent of the global monsoon region, characterized by large  DSI, is projected to remain substantially  unaltered.      
Centroid analysis shows that most of  CMIP5 projections  suggest that the monsoonal annual rainfall distribution   is  expected to change from early to late in the course of the hydrological year by the end of the twenty-first century and particularly after year 2050. This trend is  particularly evident    over Northern Africa, Southern Africa  and western Mexico,  where more than  $90 \%$ of  the models  project a delay of the rainfall centroid from a few days up to two weeks. Over the remaining monsoonal regions, there is little  inter-model agreement in terms of  centroid changes.

\end{abstract}

\section{Introduction}
\label{intro}

Climate warming induced by anthropogenic greenhouse gases (GHGs)   is expected to alter    the global  precipitation field \citep{Meehl,IPCC5}. The reason for that is twofold. First, the thermodynamic effect of   higher surface and near-to-surface atmospheric temperatures will  be to   increase    specific humidity, thus strengthening the hydrological cycle even if the atmospheric circulation were to remain fixed   \citep[][]{Allen, Soden,  Chou2}. Second,   because of the strong energetic coupling between the water cycle and   the atmosphere due to surface latent heat fluxes, radiative diabatic heating  and condensation diabatic heating, modifications  in the atmospheric water cycle will  impact, in turn,  the general circulation of the atmosphere  \citep{Alessandri, Bony}. This implies  that changes in the hydrological cycle can take place  even if specific  humidity were to remain fixed due to dynamic effects \citep[][]{Seager}. With both the thermodynamic and dynamic mechanisms at play under GHGs forcing,   a satisfactory comprehension  of the effects of global warming on precipitation at both large and small scales requires a more robust comprehension  of the response of atmospheric circulation-related fields  \citep{Shep}.

Currently, global climate models (GCMs) are our  primary tools for projecting future changes in the hydrological cycle under GHGs forcing scenarios. Although they have undergone through intense development over the last thirty years, a precise representation of the hydrological cycle has not been achieved yet, particularly at the local and continental scales \citep{Danube, Lin,Boos, Shabeh1, Sperber, Shabeh2}. This is due to  misrepresented  small-scale hydro-climatic processes \cite[e.g.][]{Sabeerali} and structural limitations as, for example, insufficient representation of complex orography \citep[e.g.][]{Palazzi, Kapnick} and uncertainty in the representation of sub-grid scale processes (e.g. clouds, gravity-wave drag, convection and boundary-layer). In spite of this, most of GCMs are successful in capturing well some  key features of the large scale precipitation field and simulate  trends such as the observed northward shift of the low-level monsoonal jet \citep{Kitoh, LeeWang} and the mean surface temperature trends \citep{IPCC5}, thus providing some confidence on their projected changes.

Most of the state-of-the-art  GCMs   agree on simulating     a precipitation reduction in the    subtropical arid and semi-arid regions and increased  rainfall in the  equatorial  and high-latitude regions  by the end of the twenty-first century under medium-to-high emission scenarios \citep[e.g.][]{Knutti2}.    If such changes  occur, they are going to have  substantial consequences for ecosystems and human activities. Wet regions may, for example,   be subject to increased flooding, whereas   semiarid areas  placed at the borders of  the subtropical subsidence regions could see  further  reductions   of rainfall and  more prolonged droughts, thus worsening their  intermittent hydro-climatic regimes \citep{Polade}.  

Apart from knowing   how the  mean  annual rainfall amount will change in the future,    a more complete description  of the modifications  of the hydroclimate   also requires to know    how other   precipitation  characteristics     will be affected by global warming: the   statistics of   intense precipitation events   \citep[e.g. frequency of events above  a certain intensity, ][]{Easterling, Goswami2, Wentz,  Coumou, Mehran, Lau, Liu}, change in the number of dry days and inter annual variability \citep{Polade}  and precipitation seasonality  \citep{Huang, Porporato, KumNat}, particularly for monsoonal regimes \citep{Seth, Kitoh, LeeWang, Turner}.  Here we will focus on these last two  aspects.

Rainfall seasonality and duration of drought events are important factors for the Mediterranean-type and monsoonal rainfall regimes.  In the semi-arid  Mediterranean-type  regions, reduced rainfall together with extended dry periods  may   exacerbate  the  risk of prolonged drought,  with potentially  very severe consequences for agriculture, ecosystems and human activities \citep[e.g.][]{Giorgi}. In the Tropics, the seasonality of rainfall is as important as the total amount of rainfall and it is the major controlling factor of the calendar of agricultural activities. Rainfed agriculture -- diffusely practiced  in Asia, Africa and Latin America --  provides food to millions of people and   depends critically   on the regularity of seasonal rainfall. Ecosystems in tropical wetlands are also extremely sensitive to the timing of  arrival of rain at the beginning of the wet season and to  the duration itself of the wet season \citep{Borchert, Eamus, Rohr, Guan}.   Substantial changes in the timing or duration of the wet season may therefore have  a  huge impact on   economies  of many countries  through its influence on  water resources.  

New  rainfall metrics have been recently introduced by  \cite{Porporato}, based on  a probabilistic interpretation of rainfall fractions  and the   concept of  relative entropy (RE) borrowed from information theory \citep[e.g.][]{Cover}. Relative entropy  is used to    quantify the diversity  between  the time series, for a given year,    of the monthly fraction of the annual precipitation  and  the   uniform precipitation sequence. The RE provides a  threshold-independent   metric to quantify  the concentration of annual rainfall and the number of dry months. 
Furthermore, it exists a  linkage between RE and the effective number of wet/dry days, and thus for arid subtropical and semiarid Mediterranean-type  regions the RE can be used as an objective  drought  index.
The probabilistic interpretation of the monthly rainfall fractions  also   allows us to estimate circular statics moments associated describing the annual distribution of rainfall and, in particular, the center of mass (centroid) of the annual precipitation, which for seasonal rainfall regimes is a measure of the timing of the wet season. 

By using these metrics, \cite{Porporato} examined   six decades of rainfall data in the tropical region of the globe, beginning from 1930,  and  found that the    timing, magnitude  and duration of tropical rainfall are  all significantly changing. 
 \cite{Pascale} applied such metrics    for characterizing  the seasonality of precipitation regimes  as represented by   observational datasets and CMIP5  GCMs over the historical period 1950-2005. 
  Here, we aim at assessing,  through  these novel  metrics, the projected    changes in the  timing, magnitude  and duration  of wet  and  dry periods over the tropical and subtropical regions     by the end of  the twenty-first century,   as simulated by twenty-four models participating   to Coupled Model Intercomparison Project phase 5 (CMIP5)  \citep{Taylor} under  the most extreme representative concentration pathway, RCP 8.5 \citep[][]{RCP}. A particular focus will be  on changes of rainfall seasonality of  six large areas --North and South America, North and South Africa, South Asian and Australia  -- belonging to the global monsoon domain \citep{Trenberth00,Ding}. The analysis  we present here  differs from previous ones  on global and regional  precipitation response to global warming in that we use a new,  coherent methodology  based on  a probabilistic interpretation of rainfall fractions. This approach  allows us to estimate circular statics moments for the rainfall timing and a threshold-independent,  information theory-based  quantity such as the RE to quantify  the concentration of annual rainfall and the number of dry months and to build an index of monsonality (DSI).
   
The structure of this  paper is the following. In Sect.~\ref{data_method} we deal with  the datasets and the indexes used for our analysis. In Sect.~\ref{results} the climatology of such indicators is  presented and discussed in the context of atmospheric general circulation.  Results on future changes of the seasonality properties of tropical rainfall are presented  in Sect.~\ref{results}    and the main findings summarized in Sect.~\ref{conclusion}.

\section{Data and method}
\label{data_method}

\subsection{Historical and twenty-first century RCP 8.5 simulations}
\label{data}

Twenty-four CMIP5 models \citep[][]{Taylor, Gui}    are taken into account and their   historical simulations  (1930-2005) and twenty-first century scenario (2006-2100) are analyzed. In Table~\ref{tab:1} the list of the models, their horizontal resolution and the names of the research institute where they were developed are shown. Further details on the CMIP5 experiment can be found online at  \url{http://www-pcmdi.llnl.gov/}. Projections available  in the CMIP5 intercomparison database  are generated by  the latest generation  of coupled climate and Earth system models. Relative to the previous intercomparison phase CMIP3 \citep[][]{Meehl},   CMIP5 models  generally have a  higher spatial horizontal resolution of the atmospheric component (nearly one-third of the models have an atmospheric resolution of about $1.5^\circ$ in latitude), more Earth System Models including biogeochemical processes \citep{Taylor} and  improved schemes for aerosol-cloud interactions \citep[e.g.][]{Wilcox}.  In spite of the huge efforts to include more complexity  and  increase  computational capacity, the improvement of CMIP5 models with respect to CMIP3 is relatively modest \citep{Knutti2}.

Simulations of the historical period (Hist) are   forced with both anthropogenic and natural forcings while those for the twenty-first century are based  on  representative concentration pathways (RCPs) \citep{Moss, Meinshausen}. We analyze the    representative concentration pathway RCP 8.5  \citep{Riahi},  in which an anthropogenic radiative forcing of $8.5$ W\,m$^{-2}$ and a mean global warming of about $4$ K  is reached in  the year 2100  with respect to the beginning of the pre-industrial period. This corresponds to  a scenario in which the  use of  fossil fuels will continue  unabated and with no mitigation.  Multiple ensemble  members for each considered scenario are available for most of the CMIP5 models selected here. In this study  we  selected just the first ensemble member (``r1i1p1'')  for each model. Furthermore, we  left   out   GCMs with serious inconsistencies in the water cycle in their control (unforced)  runs for  which     long-term annual means of evaporation minus precipitation is   larger, in absolute value,  than  $10^5$\,m$^3$\,s$^{-1}$  and  equivalent to an energy  bias larger than  $\approx1$ W\,m$^{-2}$  \citep{Liepert,Liepert2}. Models are compared over land with the GPCC \citep{GPCC,GPCC2} and CRU \citep{CRU} gridded precipitation dataset, and over the whole global surface with the GPCP dataset \citep{Xie2}.  Given the different horizontal resolutions among models themselves (Table~\ref{tab:2}) and observations ($1^\circ\times 1^\circ$), all processed data from models and observations  are linearly interpolated to a common $2.5^\circ\times 2.5^\circ$ grid  when necessary.

\subsection{Rainfall seasonality indicators} 
\label{method}

Rainfall properties  are described through the following indicators  of the magnitude, timing and duration of the wet season \citep{Porporato, Pascale}:  
\begin{itemize}
\item[1.] \emph{Annual rainfall.}
Given the  monthly precipitation sequence $r_{m,k}$ (month $m$, year $k$), the total annual rainfall during year $k$ is simply $R_{k}=\sum_{m=1}^{12} r_{m,k}$.

\item[2.] \emph{Relative entropy.}
From the   monthly precipitation fraction    
\begin{equation}
\label{prob}
p_{m,k}=r_{m,k}/R_k,
\end{equation}
the relative entropy of the $k$-th year  is defined such as 
\begin{equation}
D_k = \sum_{m=1}^{12} p_{m,k}\log_2\left( 12\,p_{m,k} \right)
\label{again}
\end{equation}
 and it quantifies how different $p_m$ is  from the monthly uniform precipitation sequence $q_m=1/12$, $m=1,\ldots,12$. The relative entropy    attains  its maximum value $\log_2 12$ when the annual rainfall is concentrated in one single month ($p_m\log_2 p_m\rightarrow 0$   is taken for $p_m=0$) and  equal to zero for a uniform precipitation sequence. As discussed in \cite{Pascale}, the RE is related to  the support of the function $p_m$ (effective number of degrees of freedom) defined by $ n^\prime_k=12\cdot 2^{-D_k} $ and interpreted as  the number of months of the wet season. 
 For monsoonal hydro-climatic regimes, characterized by unimodal monthly rainfall sequences, $n^\prime$ provides a measure of the duration of the wet season.   The number of dry months  $\tilde{n}_k$ for each hydrological year  is thus
 \begin{equation}
  \tilde{n}_k=12-n_k^\prime=12(1-2^{-D_k}).
 \label{effective}
 \end{equation}
 The definition of $\tilde{n}$ used here is different from  that used in recent studies \citep[][]{Lau, Polade} in that  it does not involve any arbitrary thresholds (e.g. $0.024$ mm/day in \cite{Lau}) in order to  define a  dry month and wet season onsets/retreats.  Although  absolute thresholds for precipitation provide very stringent tests for climate models \citep{Sperber}, their use may lead to systematic biases due to the ``dryness'' or ``wetness'' of single models, as recently shown by \cite{Sperber2}  to assess  the onset and retreat of the Asian monsoon in climate models.
  
\item[3.] \emph{Dimensionless seasonality index.} The dimensionless seasonality index (DSI)  is  defined as \citep{Porporato}
\begin{equation}
S_k=D_k\left(\frac{R_k}{R_0}\right)
\label{dsi}
\end{equation}
in which $R_0$ is a constant  scaling factor introduced in order to make the precipitation dimensionless. We choose $R_0$ as the maximum of $R$ of  observation gridded datasets over the whole period ($9701$ mm). According to definition (\ref{dsi}), $S_k$ is zero when either $R$ (completely dry location) or $D_k$ ($R_k$ distributed uniformly throughout the year) are zero and maximum ($\log_2 12$) when $R_0$ is concentrated in a single month.   In \cite{Pascale} it is shown that the DSI, together with the RE,    appears to be a useful indicator  of the  extent of the global monsoon region  (GMR) \citep{Ding, Wang, LeeWang}, which generally feature large values of RE and DSI. In the following   the GMR is defined as the area such that  $S>0.05$ and $D>0.4$.

\item[4.] \emph{Centroid.} An index  of the timing  of the wet season can be derived directly from the  definition of  $p_m$ in Eq. (\ref{prob}).  From circular statistics \citep[e.g.][]{circular2},  the   centroid  $C_k$     is defined as the argument of the complex number $z_k$ ($\textbf{i}$ imaginary unit)
\begin{equation}
\label{centro}
z_k=\sum_{m=1}^{12} p_{m,k} \exp{\{ \textbf{i}  2 \pi m/12 \} }, \quad   C_k=\textrm{arg}(z_k).
\end{equation}
The centroid  is an integral measure of the timing of the wet season and   it is related to    time of the year around which most of the annual  rainfall  distributed. Therefore    it   cannot inform us about the beginning and the end of the wet season \citep[e.g. onset and retreat for monsoons][]{Shabeh2, Sperber, Kitoh}. It has to be noted  that, although $C$ can be formally defined for any $p_m$, it is not very meaningful  for regions with low RE -- that is for regions without a clear dry season during the year. Changes of $C$,  associated with an early-to-late or late-to-early   redistribution of rainfall   between  2080-2100 and 1980-2000,  are obtained as the phase difference 
\begin{equation}
\Delta C=\textrm{arg}(\zeta/Z)
\label{delay}
\end{equation}
where     $Z=\sum_{k=1980}^{2000}  z_k\equiv |Z|\exp{\{\textbf{i}\, \textrm{arg}(Z)\} }$ and $\zeta=\sum_{k=2080}^{2100} z_k \equiv |\zeta|\exp{\{ \textbf{i}\, \textrm{arg}(\zeta)\} }$.

\end{itemize}

We compute   $p_{m,k}(x)=r_{m,k}/R_k$, $R_k(x)$,  $D_k(x)$  and  $S_k(x)$  for each  model gridpoint  $x=(\phi, \lambda)$ ($\phi$ latitude, $\lambda$  longitude) and for each   hydrological   year $k$. Differences between the climatological means  $\overline{R}(x)$, $\overline{D}(x)$, $\overline{S}(x)$        over  the  time periods  2100-2080 and 1980-2000 are calculated in order to analyze the effects of GHGs forcing on precipitation.

\section{Indexes' climatology}
\label{clim}

\begin{figure*}
\centering
\subfigure{\includegraphics[angle=-0,width=1.2\textwidth]{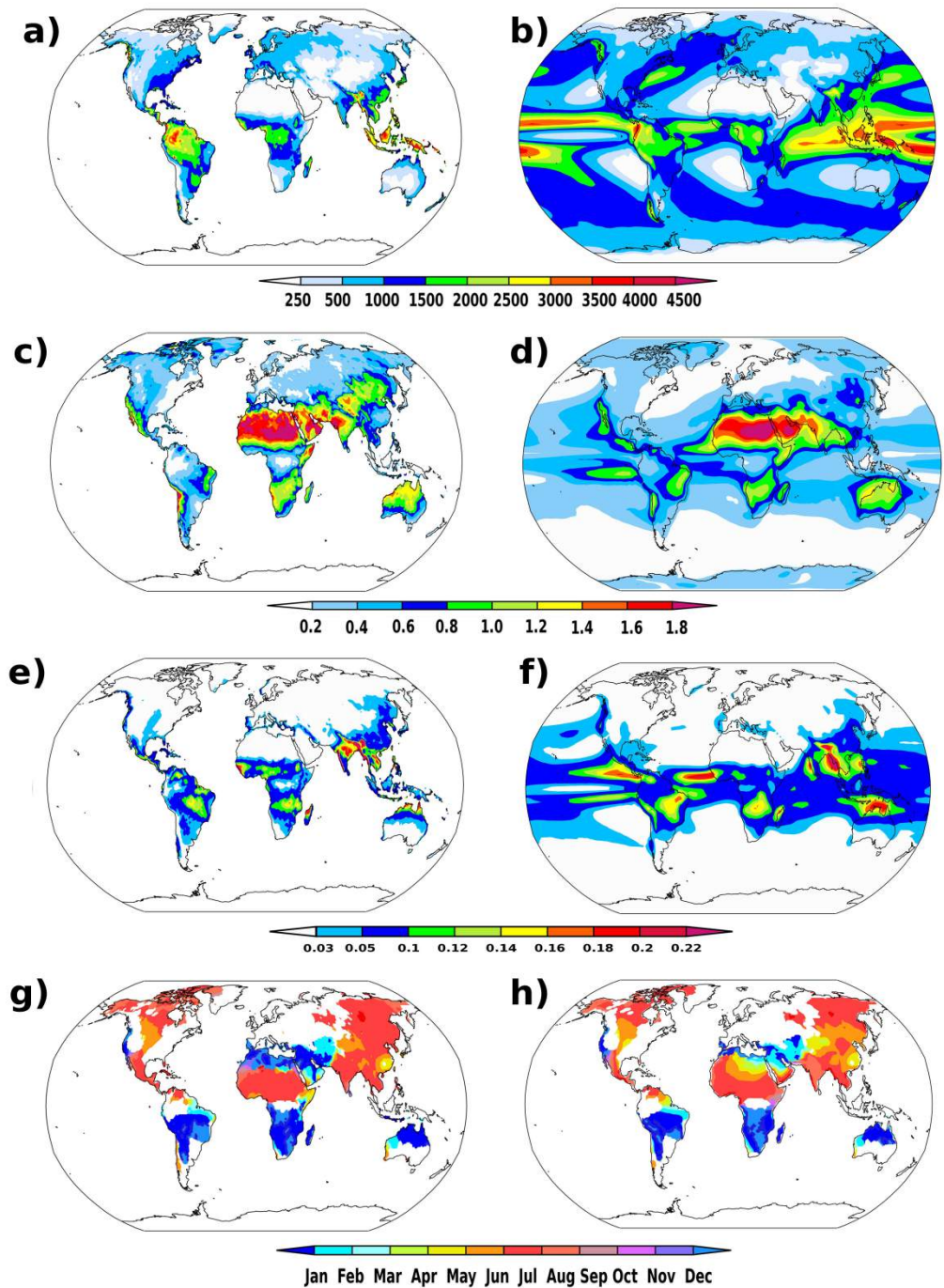} }
\caption{Climatologies (1950-2010) of: (a, b) mean annual precipitation (mm/year); (c, d) relative entropy; (e, f)  dimensionless seasonality index  and (g, h) centroid  from the GPCC precipitation dataset and CMIP5 historical simulations (MEM) respectively. Values of the centroid are shown only for those areas having $D\ge 0.3$ -- corresponding approximately to $n^\prime$ greater than 10.}
\label{climatology}      
\end{figure*}

The climatological means (1950-2100) of  annual  precipitation, relative entropy, dimensionless seasonality index and centroid  are shown  in Fig.~\ref{climatology}  for  the updated  GPCC land precipitation dataset \citep{GPCC,GPCC2} and for the CMIP5  Multi-model Ensemble Mean (MEM)\footnote{ The  MEM estimates do not necessarily give a better indication than the best performing model  \citep{Tebaldi}. Therefore we present the MEM,  together with single models output,   just for giving an indication about the behavior of most of the models used in this study.}.
Regions with the largest RE (Fig.~\ref{climatology}(c, d)) are those placed in the arid and    semi-arid subtropical subsidence  belt  between  $10^{\circ}$ and $30^{\circ}$ N/S (southern Africa, eastern Brazil, north Australia, western India, eastern Siberia, eastern Mediterranean sea,  western  Americas, Middle-East and central Asia) where the little  yearly  rainfall  is  concentrated in a few months.   

The  DSI (Fig.~\ref{climatology}(e, f)) has its largest values  over the global monsoon  region   \citep{Trenberth00, Ding, Wang} -- North American Monsoon region \citep[NAM,][]{Cook}, the South American Monsoon region \citep[SAM,][]{Braz}, the North African Monsoon region \citep[NAF,][]{NAF}, the South African Monsoon region (SAF), the South Asian Monsoon region \citep[SAS,][]{Turner} and the Australian Monsoon region \citep[AUS,][]{Austr}. Large DSI over these areas are due to    intermediate-to-high  levels of annual rainfall  and  large  relative entropy.  As observed,  models reproduce fairly realistic pattern of precipitation, RE and DSI but tend to show some systematic biases in RE     over  tropical Latin America and  underestimate it in  Western Africa and  East Asia.  \cite{Pascale} shows that     positive   RE biases in a GCM   are associated with  simulated  monthly precipitation fractions which are   too large during the wet months and too small in the months preceding the wet season whereas  negative biases are instead  due to an excess of rainfall during the dry months.

The centroid pattern  can be seen in Fig.~\ref{climatology} (g, h)) for both GPCC land observations and CMIP5 models (MEM).   The centroid clearly shows the timing of the wet season in the  monsoonal regions (e.g. July in Northern Africa or January in Northern Australia) and mediterranean climates (e.g. Middle East).  Areas of transition from monsoonal  summer precipitation regimes  to Mediterranean-type winter  precipitation regimes (e.g. Indus valley and Southwest U.S.) are characterized by a sharp transition from summer  to winter months. Agreement between observation and multimodel ensemble means  over land is reasonable, although some discrepancies are present. Over most of India,  the simulate  rainfall centroid falls in  August instead of July.  A delay is observed also over the South American monsoon region, where models tend to place the centroid in January instead of December.  Good agreement is found over the North  and South African monsoon regions,  and over Australia. 

The centroid MEM's delay relatively to GPCC observations -- as over India and South America --  is associated with pronounced  negative precipitation anomalies in the early monsoon season, as evident from  the zonally-averaged annual cycle of regional monsoon precipitation (latitude-time) shown in Fig. 2 of \cite{Seth}.  Fig.~\ref{climatology}(g, h) also shows a detailed geographical view of such  biases, which is inevitably lost in the  zonally-averaged H\"ovmoller diagrams of precipitation. For example, most of the Indian region contribute to  the negative May-July  rainfall anomaly  over India shown in \cite{Seth}, while  only the northern part of South American monsoon region is interested by precipitation anomalies in the early monsoon season. Precipitation  is instead  underestimated over North Australia  or overestimated over South Africa fairly uniformly along the whole monsoon season, resulting in no net centroid bias.

A more complicated picture emerges instead over the Southwest U.S., placed at the border between the prevalently  monsoonal-type regimes of  central-western Mexico  and the Mediterranean-type precipitation regimes of California. Here  models simulate a precipitation centroid overly shifted  towards  the  autumn.  This  inconsistency, which becomes more and more serious  as moving northwards toward California,   arises because of two model errors: the tendencies of CMIP5 models to overestimate the late-monsoon rainfall \citep[e.g.][]{Geil, Sheffield} and  the excessive intrusion southwards of winter midlatitude cyclones originating over the Pacific ocean, yielding  a positive precipitation anomaly in December-January \citep{Seth}. Precipitation over transitional areas   --  e.g. Southwest U.S, Indus basin \citep{Shabeh2}, Sahara, East Africa  --  challenges GCMs and consistent biases are visible, over these regions, in Fig.~\ref{climatology}(g, h).

\section{Twenty-first century RCP8.5 projections}
\label{results}

\subsection{Magnitude: projected changes of mean annual precipitation}
\label{MAP}

 The MEM  of  percentage precipitation difference between the  RCP8.5  and Hist  (2080-2100 minus 1980-2000) is shown in Fig.~\ref{precipitation}. As reported already by several authors \citep[][]{Ren, Knutti2, Chadwick, Kitoh} and by the IPCC5 Assessment Report \citep{IPCC5}, CMIP5 models under the RCP8.5 scenario  project an increase in the annual mean precipitation  in most of the  high-to-mid latitude regions and  in the  equatorial Pacific Ocean; subtropical subsidence zones and  Mediterranean-type regions placed at the equatorward margins of the midlatitude storm tracks        are instead expected to see a decrease in their  mean annual precipitation. Differences with respect to the previous generation of climate models (CMIP3) have been  shown to be minimal \citep{Knutti2}. 
 
 Time series  of the  of the averaged precipitation anomalies\footnote{Note that intermodel differences in terms of precipitation can be very large \citep[e.g.][]{Sperber2} and are partially hidden in showing the anomalies.} relative to 1930-2000  over the  six tropical monsoonal regions listed in Table~\ref{tab:2} are shown in Fig.~\ref{precipitation}. We use the 25th and 75th percentiles  and  the total inter-model spread  to give a  statistical description of how the various model outputs are  scattered. Inter-model spread is large particularly over the NAF, SAM, SAF and AUS regions (ratio of inter-model spread to signal), making hard  to interpret   the GCMs projections over such areas.  Over the NAM and SAS regions inter-model agreement is found to be higher, with about $90\%$ of models agreeing on a reduction of rainfall over the NAM region and an increase over the SAS region.  Comparison with  observations  reveals that  models  generally tend to underestimate natural variability, as evident from the GPCC and CRU lines often wiggling outside  the total inter-model spread. This is particularly  evident  for the NAF region, for which the signal associated with  the 1950-1980 Sahel drought  \citep{Dai}  is well outside the total inter-model spread.  As highlighted by \cite{Knutti2},  model agreement is fairly low in the Tropics and generally better at high latitudes. In particular, changes are not robust over SAM, NAF and AUS, as reflected  also  by the time series in Fig.~\ref{precipitation}.

\subsection{ Duration and number of wet months: projected changes in relative entropy}
\label{RE}

We   assess here the changes in rainfall monthly fractions and number of dry months analyzing the RE metric.   The MEM  of the projected changes in  RE is shown in  Fig.~\ref{entropy} together with the  time series  of the area-averaged    RE anomalies over the  six monsoonal regions. Single-model projections of RE anomalies are instead shown in Fig.~\ref{pannel1}.  

But what do  projected RE changes practically mean in terms of precipitation regimes?  Changes in RE  imply  changes in the number of wet and dry months, $n^\prime$ and $\tilde{n}$  (Eq.~\ref{effective}). In particular,  areas where RE is projected to increase   will see a reduction of   $n^\prime$ and thus an increase in the number of dry months. 

Over monsoonal regions  \citep{Ding},  rainfall is concentrated during  the (local) summer period and the dry season coincides with the (local) winter. Here a   reduction of $\tilde{n}$  --  i.e. an increase of RE  --  would mean a reduction of the duration of the wet season. A projected increase of RE might  be associated  either with  an  increase of  the percentage of local summer rainfall  over the total annual rainfall \citep{LeeWang}  or with a shorter duration of the wet season without any reduction of the summer rain  (or with  both of them).  If more rainfall occurs during the middle of the monsoonal season, the associated rainfall fractions $p_m$ will be even larger than  those associate with the pre-monsoonal and post-monsoonal months. In this sense, winter  months become  ``drier'' even if there is no reduction at all during them.  Let us stress here that changes in wet season duration  based on the RE do no necessarily agree with onset/retreat estimates based on absolute thresholds \citep{Sperber}, although a better agreement exists with relative threshold approaches such as in \cite{Sperber2}.

 For arid and semiarid   Mediterranean-type and subtropical subsidence regions, where there is not a regular  wet season and rainfall is typically erratic during the year,   the RE still  quantifies $n^\prime$ and $\tilde{n}$, thus  providing  threshold-independent, objective  drought index. 
Figure~\ref{duration} shows the MEM differences in $\tilde{n}$ between 2080-2100 and 1980-2000 under the RCP8.5 scenario. Consistently with the RE anomalies  shown in Fig.~\ref{entropy}, more than 90$\%$ of the CMIP5 models predict  a notable  increase  of $\tilde{n}$ ($\Delta \tilde{n}$ up to one month) in the  Mediterranean area, southern Europe, western Mexico-Southwest U.S.,  southern Africa, eastern Brazil and South Australia. Agreement with Fig.2 of \cite{Polade}, who used instead a threshold-dependent method to define a dry day,  is remarkable.

 \subsubsection{RE changes in high-latitude and Mediterranean-type regions}
 
 Overall,   CMIP5 RCP8.5 projections suggest a   RE  decrease  by the end of the twenty-first century for high-latitude regions north of $50^\circ$ N and south of $50^\circ$ S. This trend is particularly strong over eastern Siberia, where  moister conditions are predicted by CMIP5 models  (Fig.~\ref{precipitation}). The projected increase in mean annual rainfall and  decrease in RE  indicates a larger  number of wet days rather than by changes in precipitation intensity.

A quite coherent pattern of RE increase can be noted  along the equatorward flank of the storm tracks in both hemispheres around $40^\circ$. Such RE changes  are likely to be associated with the projected poleward shift of the storm tracks \citep{Bengtsson, LucariniRagone} and consequent transition towards more seasonal and erratic precipitation regimes. 

Most of GCMs consistently show a strong increase of the mean annual RE 
 ($\approx 0.2$) by the end of the twenty-first century over the Mediterranean-type regions such as northern Africa, southern Europe, southern Australia and   semi-arid regions such as northwestern Mexico. Such a large projected change in RE is related to an increase of number of dry days, $n^\prime$ (Fig.~\ref{duration}). 
In particular,  southern and southeastern Mediterranean coastal regions feature a very dry June-September period  with precipitation mostly concentrated between  November and February  \citep{Mehta}, and therefore relatively high values of RE (Fig.~\ref{climatology}(c)).  Drying of the Mediterranean region is predicted by CMIP5 models and has been documented in several studies \citep[e.g.][]{Kelley, Baker}.  The remarkable drop in $n^\prime$, which  is most likely associated with a lengthening of the dry spells, and conditions of  reduced mean annual precipitation  are expected to have a high impact on local water resources, agriculture and ecosystems.  Further research is needed for a more precise explanation of the projected modifications  of the rainfall seasonality   and of the circulation mechanisms involved in it.
 
The Mediterranean pattern of positive $\Delta \tilde{n}$ visible in Fig.~\ref{duration} extends northwards  into northwestern Europe and  the British Isles, for which no mean annual rainfall reduction is predicted  by CMIP5 models.   This gives an indications that dry period may become more prolonged, and therefore drought more severe, also in areas like northwestern Europe generally featuring relatively high  annual rainfall amount and no pronounced precipitation seasonality.  Let us also  note that even areas of high  agreement amongst  models might not imply more confidence about the correctness of the  projections, since  common errors or deficiencies in physics  parameterizations  may provide  false confidence in the future projections' robustness.


\subsubsection{RE changes in tropical and monsoonal regions}

 Tropical and subtropical regions instead show a more diverse picture. Consistent RE increases are  projected over  western Indonesia, eastern Brazil and southern tropical Africa.  However GCMs projections over central and most of northern Africa, South Asia and a large part of Latin America are less coherent (Fig.~\ref{pannel1}).  
 Nearly $90\%$ of the CMIP5  models   robustly project  an increase of RE over three out of the six monsoon regions selected for this study:  NAM,  SAF and SAM. 
 
  In particular, over the SAF area changes in annual rainfall are small and insignificant, whereas  most models project consistent RE anomalies relative to the 1980-2000 period of about   0.1-0.2.  Similarly,  an RE increase is foreseen also over  tropical Latin America and specially over its eastern area with little and insignificant changes in the mean total annual rainfall.   This suggest that the (almost) same amount of annual rainfall  is projected to be distributed differently, on average, by the end of the 21st century  and to be more concentrated during the wet season, exacerbating the droughts of the  dry winter months and increasing flood risks.  

CMIP5 models suggest   a reduction of precipitation over Southwest U.S. and western  Mexico \citep{Seager2}, although large differences exist in what these models predict (Fig.~\ref{precipitation}). However their agreement is much higher in terms of the sign of the RE, indicating an increase of about $0.1$, (with a few models, e.g. MPI-ESM-LR  or IPSL-CM5A-MR  Fig.~\ref{pannel1}, up to 0.4) and corresponding to a median increase of $\tilde{n} $ of almost one month (Fig.~\ref{duration}). Such an increase in the number of dry months is coherent with  with the marked rainfall reduction (up to $40 \%$) between  December and May   \citep{Maloney}, leading to a much drier winter-spring period.
  
  The  positive RE   projection   over the SAS region is much less robust across models  ($\approx 70\%$ of the models), with single models showing very different responses (e.g. MRI-CGCM3 and MPI-ESM-LR featuring opposite behaviors, Fig.~\ref{pannel1}).
  Positive  RE anomalies are consistent with an increase of monsoonal precipitation during summer, that is with the annual range of precipitation (summer minus winter) \citep{LeeWang}.

 We have instead no confidence in the positive trends over AUS region  and negative trend over  the NAF region, since  models show a large spread and opposite signs  in their RE anomaly response during the twenty-first century. The almost zero trend over the NAF region seems to be  due to a cancellation between  a persistent positive trend in the western part of the NAF domain (Western Africa) and a negative trend in the eastern part.  
   Let us finally note  that model biases in some regions may decrease confidence in RE projections, including changes of the RE over the SAM region and over eastern Asia that exhibit projected RE changes in the same sense as historical biases \citep{Pascale}

 
\subsection{Global monsoon regions}
\label{DSI}

 \cite{Pascale}  showed that the DSI, together with the RE,   is  a useful indicator  of the extent of the global monsoon region  (GMR)\citep{Ding, Wang, Kitoh,LeeWang}. Rainfall in  monsoonal regions is in fact  concentrated in typically five-six months, from late (local) spring to early (local) autumn  -- with annual accumulation usually exceeding $500$ mm/yr. Therefore, the GMR  usually features large values of both  RE and DSI.  The DSI combines intensity of rainfall and seasonality, thus excluding from the GMR dry areas with erratic rainfall regimes -- e.g.  Sahara -- which are also characterized by high values of  mean annual RE (Fig.~\ref{climatology} (c, e)).
 
  In the following   the GMR is defined as the area over which we have   $\overline{S}>0.05$ and $\overline{D}>0.4$.  In Fig.~\ref{domains} we can see a comparison of the GMR as defined through the RE/DSI criterion and as defined by the \cite{Wang}'s criterion based on the annual range of precipitation (ARP) (details in the caption).  The two methods compare  well over land (green and blued shading in Fig.~\ref{domains}), although they show some difference over oceans. 
  Note that  the DSI/RE criterion also identifies  some small extratropical  areas such as the West U.S. coast, southern Andes and the southwestern part of the Iberian peninsula.  These areas,  featuring considerable rainfall seasonality,   clearly do not belong to the GMR.
  

Projected changes in the GMR domain  by the end of the twenty-first century (2080-2100)  are also shown in Fig.~\ref{domains}.  According to the RE/DSI criterion, the limits of the GMR are not expected to change substantially over Africa, Asia, Australia and Americas, while small extensions are appreciable over the Indian and Pacific Ocean. These results thus suggest that this spatial domain is not going to have important alterations, at least on the large scale. 

Our results, obtained with  the novel metrics  RE and DSI,  thus support previous  findings  about the unchanged GMR under GHGs forcings based on the ARP indicator  \citep{LeeWang, Kitoh}. While differences between
the two criteria appear to be minor, they may still be relevant for assessing the robustness of future changes on more limited areas.
Moreover, it will be important to re-focus this  analysis on continental and regional scales using the output of higher resolution models to have  a more refined assessment of such changes, which indeed may very relevant for local communities. 

\subsection{Timing:  projected changes of the centroid}
\label{time}

For regions where the annual rainfall distribution is clearly  concentrated in a  wet season (e.g. monsoons or Mediterranean-type), $\Delta C>0$   is  due to   a rainfall redistribution from early to late  during the year while  $\Delta C <0$ is indicative of a redistribution of rainfall from late to early in the year.     This information  is  not necessarily equivalent, however,  to a statement of    delayed or moved-up   monsoon onset  (or retreat).  Onset and retreat   are in fact   defined as the day (or the pentad) rainfall attains a certain  absolute \citep{Kitoh, Sperber} or relative \citep{Shabeh2, Sperber2} precipitation threshold and thus depend from a very localized (in time) event.  


The MEM of the centroid changes   (2080-2100 minus 1980-2000) are shown in Fig.~\ref{delay}   together with time series (1930-2100)  of   the centroid anomalies area-averaged over the  six monsoonal regions defined in Table~\ref{tab:2}. 
Performances of the CMIP5 models contributing to the MEM  (Table~\ref{tab:1})  are shown in Fig. ~\ref{pannel2}. The  pattern of projected changes  is very heterogenous with large inter-model spread.  However a robust signal emerges  over tropical lands between $5^\circ$ and $20^\circ$ in both hemispheres, for which  most of models project a      positive  $\Delta C$. Patterns  shown in Fig.~\ref{delay}  compare relatively  well with that of   \cite{Dwyer}, who estimated  delays of the annual harmonic  of the Fourier transform of rainfall data  and through   EOFs analysis.  

In particular,  more than $90 \%$ of the GCMs indicate that the rainfall distribution associated with the African monsoons  will shift from early to late during the wet season.   Rainfall  is  projected to be redistributed from early to late in the NAF region and in East Africa, with  delays ranging from   three days  in the southern part of its domain    up to fifteen   days in the Sahel \citep{Biasutti}.   A closer look at the NAF region  reveals that        the median of the area-averaged centroid anomalies will   exceed the range of inter-model (Hist)  and observational spread    of $\pm 3$ days   after year 2030.

 The SAF monsoon season   is also projected to be delayed by the end of the twenty-first century with high inter-model agreement (more than $90 \%$ of the GCMs considered in this study), although the delay is less  marked  -- three to six days -- than  its northern counterpart, with the    median    exceeding  the   inter-model range (during Hist) roughly after year 2050.  Our results,  indicating a delay and a shortening of the wet season in this region,  are in agreement with those of \cite{Shongwave}, obtained with a different methodology and from CMIP3 models.
 
    A moderate delay of two-three days is visible also over the SAM and SAS regions, although less significant. This late shift of the centroid   is   probably  associated with a marked later retreat of the SAS monsoon discussed by  \cite{Kitoh}. 
 Over the SAS area, this signal is localized mostly over the east Arabian sea and west Indian coast  with high inter-model agreement.  Over the rest of the Indian region most of models still predict a delay, although the  signal is not as robust as over west India, with a larger    spread across models (Fig.~\ref{pannel2}) and with some (e.g. bcc-csm-1 or HadGEM2) predicting a late-to-early redistribution of rainfall.

A clear  change over the NAM  monsoonal region does not emerge from the analysis of centroid changes. There is a large disagreement  among models, with about half of them indicating a delay and the other half an advance (note the multi-model median  remaining unchanged) and with  models projections featuring a very large spread ($\pm 10$ days).     \cite{Kitoh}'s analysis, based instead on        the determination of the monsoon onset and retreat, also  found low agreement among models in the change of the onset ad retreat for the NAM monsoon, with no clear indication in a substantial delay.   Nevertheless,   more than $90 \%$ of  models  agree on a positive change of the centroid  just south of  of the North American monsoon region  at about $15^\circ$ N, which is an area  not included in our definition of the NAM region (Table~\ref{tab:2}).   This is consistent with the reduced Mexico and Caribbean precipitation in June and July  and increased  rainfall  in September and October \citep{Cook, Maloney}. If this region is taken into account, RCP8.5 simulations indicate a delay for  NAM wet season too, as discussed by previous authors \citep{Grantz}. 

No substantial agreement among models is found  over  the AUS region, where roughly half of models suggest a delayed wet season and the other half an advanced we season, in agreement with \cite{Seth}. \cite{Kitoh} analysis of onset and retreat  dates changes suggests that this is due to an earlier onset and later retreat.    As for the other seasonality indicators, models projections are difficult to interpret reliably over this region because of the large inter-model uncertainty. 
  


\section{Discussion and conclusions}
\label{conclusion}

We have examined the projected changes of rainfall seasonality and dry spell durations  by the end of the twenty-first century under a high GHG emission  scenario. The diagnostics  employed  in this study have been  used to have observational  evidence  of changes of rainfall seasonality in the Tropics occurred during the last century  \citep{Porporato} and  to assess  CMIP5 historical simulations \citep{Pascale}.   This recently introduced  approach  relies on  a probabilistic interpretation of rainfall fractions that   allows us to coherently estimate circular statistics moments for the rainfall timing (e.g. the centroid) and  threshold-independent,  information theory-based  quantities (RE and DSI) to quantify  the timing and time concentration of annual rainfall and the number of dry and wet months. Here, for the first time, we have  applied  it to twenty-first century RCP8.5 scenario  in order to assess how GHGs forcing is going to impact these major features of the  large-scale precipitation. 

In spite of model uncertainty and inter-model disagreement  affecting the CMIP5  projections for  precipitation, some coherent changes of rainfall seasonality in the Tropics have been identified with our methodology. Responses  are quite different for each single monsoonal region  (NAM, SAM, NAF, SAF, SAS, AUS) taken into account in our study. This   indicates   that the dynamical and thermodynamical mechanisms at play due to GHGs forcing are   diverse for each monsoon regions \citep[e.g.][]{Nie, Cherchi}.

\begin{figure*}
\centering
\subfigure{ \includegraphics[angle=0,width=0.99\textwidth]{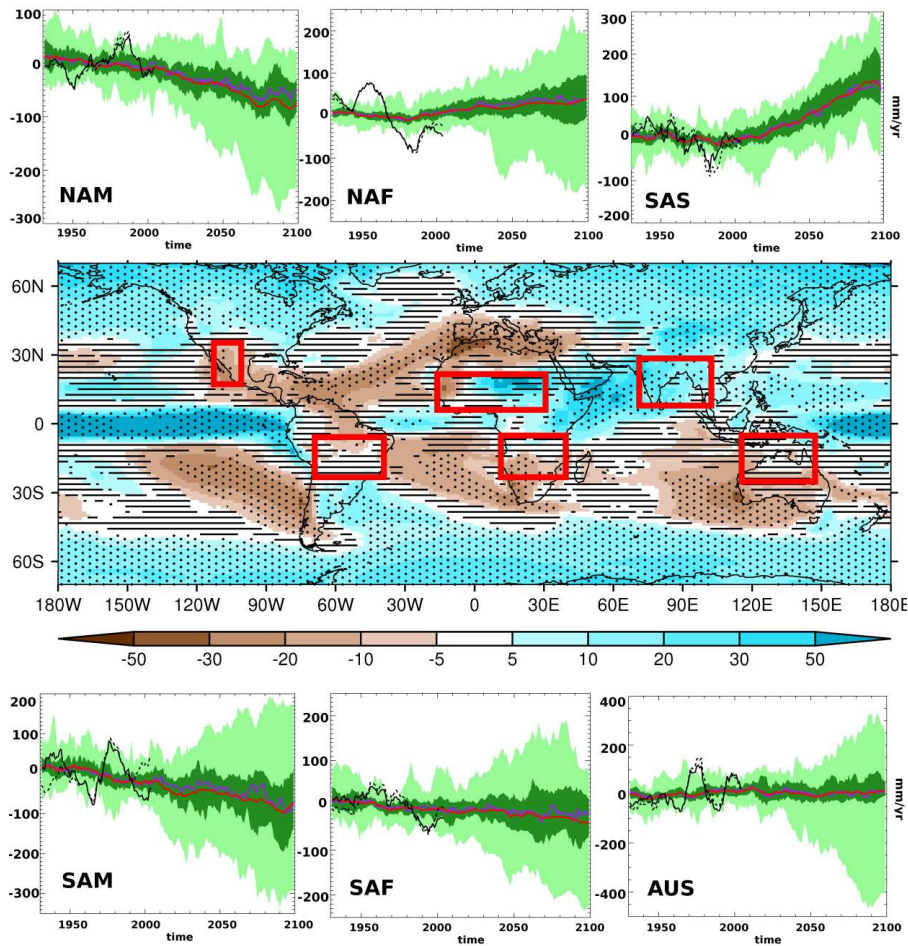} }
\caption{Relative  precipitation changes ($\%$) between 2080-2100 and 1980-2000 and    time series of  area-averaged    annual precipitation  anomalies (mm/yr) over such regions  (10-year running means) relative to the period 1930-2000. Red rectangles denote the six monsoonal  regions (NAM, SAM, NAF, SAF, SAS, AUS)  defined in  Table~\ref{tab:2}.   Areas where  at least $90 \%$ of the models agree on the sign of the change are marked with stippling  while  areas where less than $66 \%$ of the models agree in sign are marked with horizontal hatching.  In the time-series, purple line denoted the median, the light green shading the total inter-model spread, the deep green shading   the 25th-75th percentiles interval  and the  red line the MEM  of the annual mean precipitation among the 24 models. Results from precipitation gridded datasets  GPCC \citep{GPCC2,GPCC}, black solid line,  and  CRU \citep{CRU}, black dashed line,  are also shown for comparison over the period 1930-2005.}
\label{precipitation}       
\end{figure*}
\begin{figure*}
\centering
 \subfigure{ \includegraphics[angle=0,width=0.99\textwidth]{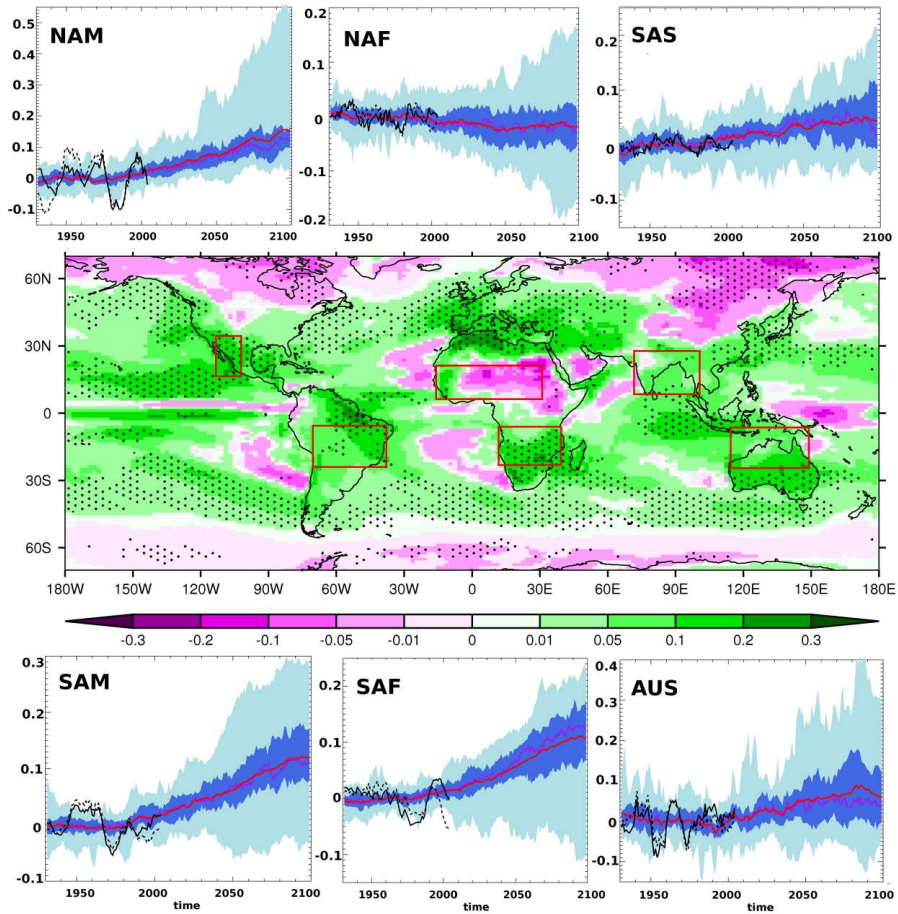} }
\caption{RE  changes  between 2080-2100 and 1980-2000 and time series (10-year running means) of the area-averaged  RE  anomalies relative to the period 1930-2000 over the NAM, SAM, NAF, SAF, SAS and AUS regions. Areas where  at least $90 \%$ of the models agree on the sign of the change are marked with stippling. In the time-series, the purple line denoted the median, the light blue shading the total inter-model spread,  the deep blue shading   the 25th-75th percentiles interval  and the red line the MEM  of the annual mean precipitation among the 24 models. Results from precipitation gridded datasets  GPCC, black solid line,  and  CRU, black dashed line,  are also shown for comparison over the period 1930-2005.
 }
\label{entropy}       
\end{figure*}
\begin{figure*}
\centering
\subfigure{\includegraphics[angle=-0,width=1.4\textwidth]{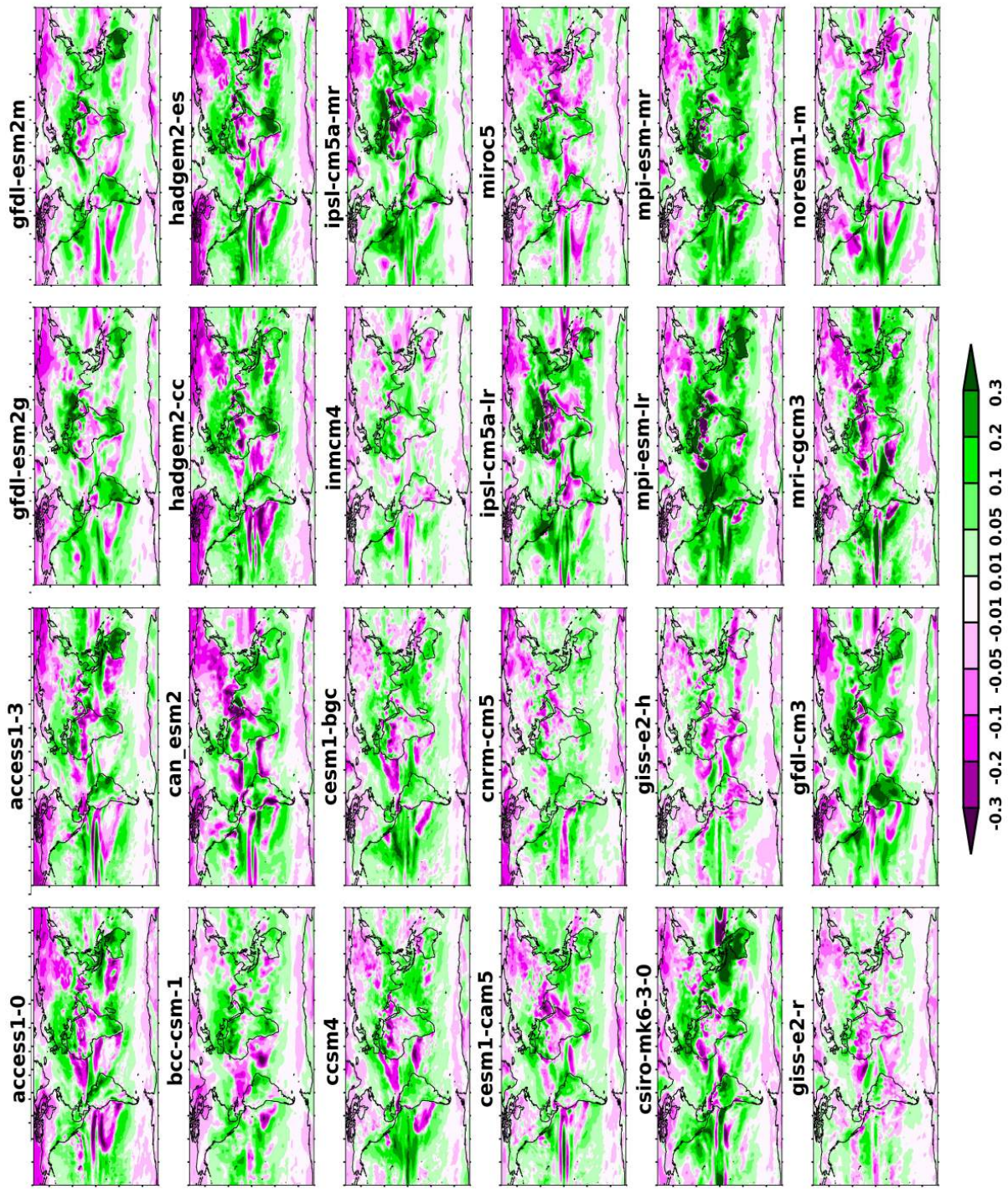} }
\caption{ Relative entropy   differences between  2080-2010 and 1980-2000  for CMIP5 models listed in Tab.~\ref{tab:1}}
\label{pannel1}      
\end{figure*}

 Most of the examined  models  indicate a significant increase of RE by the end of the twenty-first century  in areas featuring arid seasonal precipitation regimes  such as tropical Southern Africa and Western Mexico  and for wetter monsoonal regions like tropical Latin America (Fig.~\ref{entropy}). According to models projections, these regions will also experience either a slight  reduction   or no significant changes in their annual rainfall amounts (Fig.~\ref{precipitation}). The implications of such projected changes in RE may be  that   drier and more prolonged dry seasons will be more likely in the SAF, SAM and NAM regions, with increased probability of drought as well as flood. This result agree fairly well with analysis of rainfall ``monsoonality'' projected changes for less extreme   scenarios \citep[RCP4.5, ][]{LeeWang}. 

Although $90 \%$ of the examined  models   agree on increased annual totals of rainfall  over the South Asian monsoon region \citep[e.g.][]{Turner}, there is not the same inter-model agreement in projected  RE changes and therefore in the projected changes of monthly rainfall fractions.  Over North Australia instead,   high inter-model disagreement for both RE and mean annual rainfall projections makes it  impossible to have high  confidence in model predictions for the twenty-first century over this region.

The  extent of the global monsoon region \citep{Ding, Wang}, as characterized by  RE$>0.4$ and DSI$>0.05$, is projected to remain substantially  unaltered by the end of the twenty-first century, despite the considerable projected changes in timing, duration and intensity. This result further supports previous results obtained with different criterions \citep{LeeWang, Kitoh}.

Concerning the rainfall timing, we have assessed through a centroid analysis how consistent is amongst CMIP5 models the projected time shift of the wet season in the Tropics under GHGs forcing \citep{Seth, Cook, Sperber2}.  We find that   such a signal is particularly  consistent across models  over the monsoonal regions of North and South Africa with  22 out of 24 of the examined models predicting a delay and   less consistent  over  South and North America monsoon regions, where only 17 out of 24 models agree on a forward shift. No coherent  centroid delay is projected over North Australia and Asia, where models show very contradictory responses, thus making projections hard to be interpreted  over these regions     (Fig.~\ref{pannel2}). 

 The physical mechanisms behind these responses, and why they seem to occur  more strongly  in certain monsoonal regions than others,  is not yet fully understood.  While several possible links between    the forward redistribution of  monsoonal rainfall and global warming         have been studied   -- spring  inhibition (autumn enhancement)  of convective  activity   due to increased (decreased) convective stability  \citep[``spring convective barrier'', ][]{Seth2};  change in the annual cycle of SSTs   \citep{ Dwyer}; poleward shift of midlatitude storm tracks \citep{Scheff}; high-latitude SST phase delay owing to sea ice reduction impacting the tropics \citep{Biasutti, Dwyer2} -- the ultimate origin for such large scale changes of tropical rainfall seasonality are still unclear.  Table~\ref{tab:2} summarizes the findings of our analysis for the six monsoonal areas.

\begin{figure*}
\centering
 \subfigure{ \includegraphics[trim = 0mm  0mm 0mm  0mm,clip,angle=-0,width=0.99\textwidth]{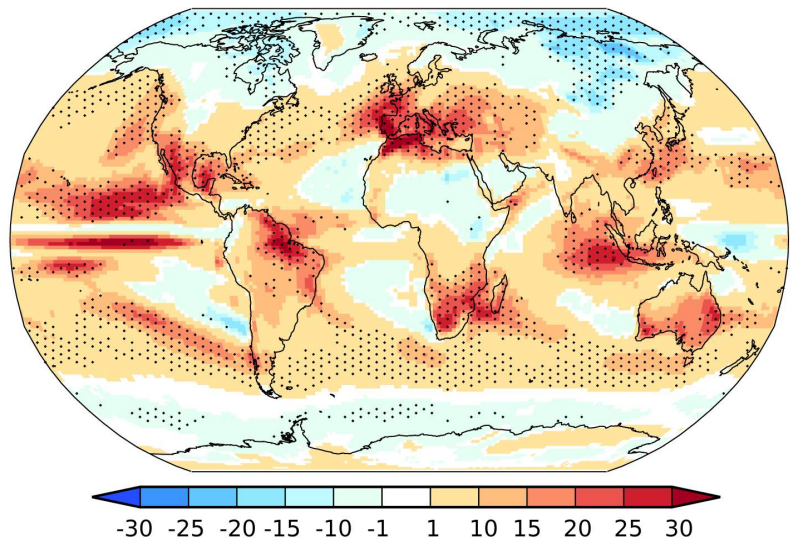} }
\caption{CMIP5 multi-model ensemble average mean change of dry days by  2080-2100, relative to  1980-2000 (days/year) under the RCP8.5 scenario. Areas where  there is high inter-model agreement on the sign of the change (at least $90 \%$ of the models)  are marked with stippling.  }
\label{duration}       
\end{figure*}
\begin{figure*}
\centering
\subfigure{\includegraphics[trim = 0mm  20mm 1mm  10mm,clip,angle=-0,width=0.99\textwidth]{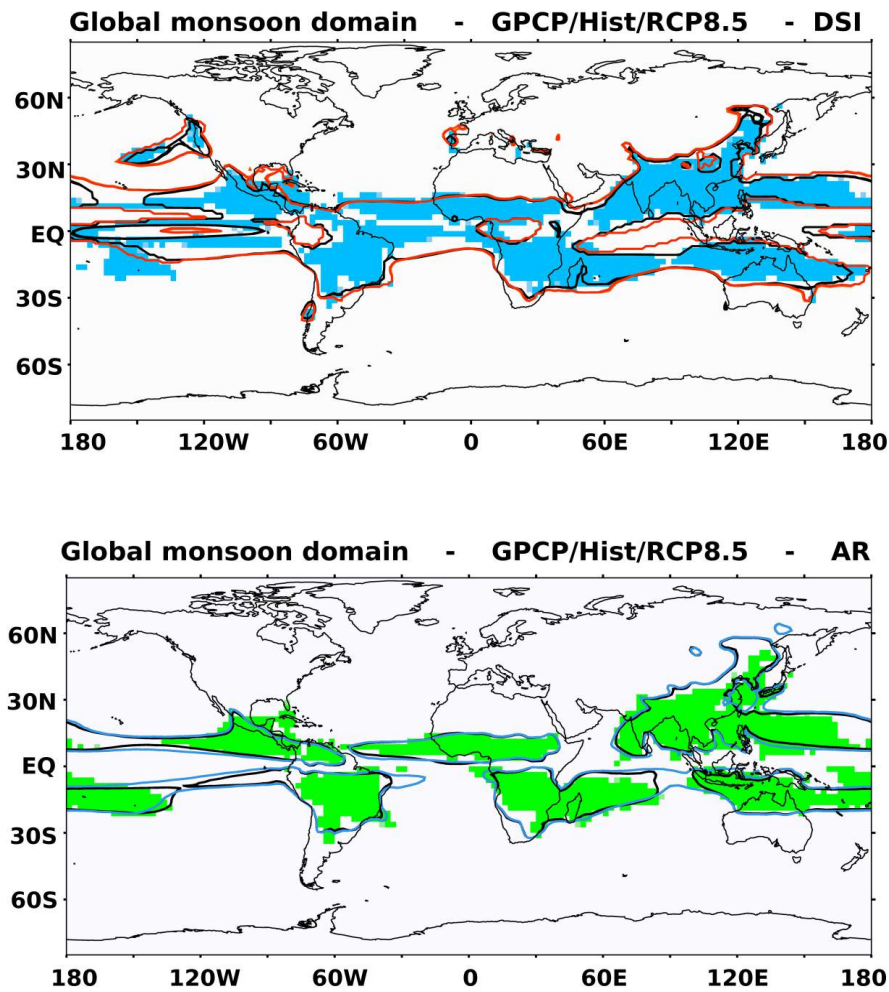} }
\caption{Observed Global Monsoon Region  (GPCP dataset) as defined by (a) the  DSI/RE metrics (blue shade) and  (b) the ARP exceeding  $2.5$ mm\,day$^{-1}$ \citep{Wang} (green shade) over the period 1979-2010. The  simulated  GMR over the historical period 1979-2010 is denoted by a black line for both methods wheres  that over the 2080-2100 period (RCP8.5) by a red line (DSI/RE) and blue line (ARP). The  ARP  is defined as the   local summer  minus winter precipitation rate, i.e.     the   MJJAS minus NDJFM precipitation rate in the northern hemisphere and  NDJFM minus  MJJAS in the southern hemisphere }
\label{domains}       
\end{figure*}

Regarding extratropical regions, projected positive changes  of  RE  in subtropical semi-arid and Mediterranean-type regions are instead associated with an increase of the number of dry days in a year.  Therefore more severe dry periods and floods  may occur in these regions, according to CMIP5 projections. This  tendency is particularly strong in southern Europe and along Mediterranean coasts, south Australia, south Africa, Central  America and eastern Amazon (Fig.~\ref{duration}). Over these regions, the overall drop in  mean annual precipitation, fairly robust across models,   has been shown to be due   the  dry-day increase    rather than  changes in precipitation intensity \citep{Polade}. Interestingly, we note that  in areas like Southwest U.S., southern Australia and western Europe the projected increase in dry days is consistent across models also in areas where no negative annual precipitation trend exist. This suggests a  redistribution of the total annual rainfall into a smaller number of days. Therefore a greater precipitation intensity may be expected  on wet days, with related enhanced flood risk and problems with water management.  Our results are in good agreement with other studies that used a different threshold-base methodology in order to define a dry day or a dry month \citep{Lau, Polade}. More research is  needed  in the future to better characterize changes in the frequency of dry-day occurrence, their spatial patterns and time  variability,  particularly on the regional scale. 


\begin{acknowledgements}
The authors  acknowledge the World Climate Research ProgrammeÕs Working Group on Coupled Modeling, which is responsible for CMIP,  and 
 the NOAA, Boulder, Colorado, USA,  for providing from their Web site  the   GPCC precipitation data.   SP, VL  and SH wish to acknowledge the financial support provided by the ERC-Starting Investigator Grant NAMASTE (Grant no. 257106) and by the CliSAP/Cluster of excellence in  the Integrated Climate System Analysis and Prediction.  AP gratefully acknowledges NSF Grants: NSF FESD 1338694, 
CBET 1033467,  EAR 1331846,  EAR 1316258 as well as the US DOE through the Office of Biological and Environmental Research, Terrestrial Carbon Processes program (DE-SC0006967), the Agriculture and Food Research Initiative from the USDA National Institute of Food and Agriculture (2011-67003-30222). XF  acknowledges funding from the NSF Graduate Research Fellowship Program.
\end{acknowledgements}

\clearpage


\begin{table}[htdp]
\caption{  List of the CMIP5 models used for this study.  \label{tab:1}}
\begin{center}
\footnotesize
\begin{tabular}{ l l l l c}
\hline
\toprule
Number & Model  name      & Modelling Centre  & Country & AGCM resolution (lon$\times$lat) \\ 
\hline
1& ACCESS1.0       & CAWCR $^\textrm{\textbf{a}}$&Australia                   & 192$\times$145/L38 \\
2&  ACCESS1.3                   & CAWCR &Australia      &   192$\times$145/L38  \\
3 & BCC-CSM1.1   & BCC $^\textrm{\textbf{b}}$&China                                &128$\times$64/L26  \\
4&  CanESM2         & CCCMA $^\textrm{\textbf{c}}$ & Canada                           & 128$\times$64/L35                        \\
5& CCSM4                        &  NCAR $^\textrm{\textbf{d}}$ &USA               &  288$\times$192/L26  \\
 6& CESM1-BGC               &NCAR&USA                  & 288$\times$192/L26  \\
7 & CESM1-CAM5                   & NCAR&USA          & 288$\times$192/L30  \\
 8& CNRM-CM5      &  CNRM/CERFACS $^\textrm{\textbf{e}}$&France            &     256$\times$128/L31  \\
 9& CSIRO-Mk3.6.0    &   CSIRO/QCCCE $^\textrm{\textbf{f}}$& Australia       & 192$\times$96/L18    \\
 10& GISS-E2-H        &  GISS $^\textrm{\textbf{g}}$& USA                                  &  144$\times$90/L40   \\
  11& GISS-E2-R        &  GISS & USA                                  &  144$\times$90/L40   \\
 12& GFDL-CM3                           &GFDL $^\textrm{\textbf{h}}$&USA             &144$\times$90/L48    \\
 13& GFDL-ESM2G                  &GFDL&USA                &144$\times$90/L24     \\
 14& GFDL-ESM2M                   & GFDL&USA             &144$\times$90/L24     \\
 15& HadGEM2-CC    &MOHC $^\textrm{\textbf{i}}$&UK                                   &  192$\times$145/L60          \\
 16& HadGEM2-ES    &MOHC &UK                                   &  192$\times$145/L38     \\
 17& INMCM4             & INM $^\textrm{\textbf{j}}$&Russia                                & 180$\times$120/L21     \\
 18& IPSL-CM5A-LR               & IPSL $^\textrm{\textbf{k}}$&France            &96$\times$95/L39    \\
 19& IPSL-CM5A-MR              &IPSL &France             &96$\times$95/L19     \\
20&  MIROC5            & MIROC $^\textrm{\textbf{l}}$ &Japan                            &256$\times$128/L40     \\
21& MPI-ESM-MR    &MPI-M $^\textrm{\textbf{m}}$ &Germany                        &192$\times$96/L95    \\
22& MPI-ESM-LR     & MPI-M &Germany                      &192$\times$96/L47      \\
23& MRI-CGCM3     & MRI $^\textrm{\textbf{n}}$ & Japan                             &  $320\times 160$/L48   \\
24& NorESM1-M      & NCC $^\textrm{\textbf{o}}$ &Norway                           & 144$\times$96/L26   \\
\bottomrule
 \hline
 \multicolumn{5}{l}{%
  \begin{minipage}{15cm}%
    \tiny \quad \\ \\\textbf{a} Centre for Australian Weather and Climate Research; \textbf{b}  Beijing Climate Centre, China Meteorological Administration; \textbf{c} Canadian Centre for Climate Modelling and Analysis; \textbf{d} National Center for Atmospheric Research; \textbf{e} Centre National de Recherchers Meteorologiques/Centre Europeen de Recherche et Formation Avancees en Calcul Scientifique; \textbf{f}  Commonwealth Scientific and Industrial Research Organization/Queensland Climate Change Centre of Excellence; \textbf{g} NASA Goddard Institute for Space Studies; \textbf{h} NOAA Geophysical Fluid Dynamics Laboratory; \textbf{i} Met Office Hadley Centre; \textbf{j} Institute for Numerical Mathematics; \textbf{k} Institute Pierre-Simon Laplace; \textbf{l}  Atmosphere and Ocean Research Institute (The University of Tokyo), National Institute for Environmental Studies, and Japan Agency for Marine-Earth Science and Technology; \textbf{m} Max Planck Institute f\"ur Meteorologie; \textbf{n}  Meteorological Research Institute  ; \textbf{o} Norwegian Climate Centre%
  \end{minipage}%
}\\
\end{tabular}
\end{center}
\end{table}

\begin{table}[htdp]
\caption{  Coordinates boundaries (east, south, west, north) of the study areas considered in this paper and shown in Fig.~\ref{precipitation}. The range of values  of the indicators is defined by the 10th-90th percentile interval (i.e. $90 \%$ of the models) of the area-averaged quantities at the end of the twenty-first century and their median (\textbf{bold} character).   \label{tab:2} }
\begin{center}
\begin{tabular}{ l  c  c  c  c | r  r r}
\hline
\toprule
Area & E       & S & W & N & $\Delta R$ (mm/yr)& $\Delta D$   & $\Delta C$ (days)\\
\hline
SAF    &  $10^\circ$ E      &  $20^\circ$S    &  $40^\circ$ W     &  $5^\circ$ S       & $[-140,\textbf{-20},+70]$   & $[0.02,\textbf{0.13},0.22]$ &    $ [1, \textbf{4}, 8]$   \\
NAF   &   $15^\circ$ W     & $5^\circ$ S      &  $30^\circ$ E      & $20^\circ$ N      &$[-40,\textbf{40}, 150]$  &$[-0.08,\textbf{-0.02},0.11]$ &   $ [2, \textbf{9}, 15]$  \\
SAS   &   $70^\circ$ E      & $5^\circ$ N      &  $100^\circ$  E   & $25^\circ$ N      & $[30,\textbf{140}, 240]$ & $[-0.02,\textbf{0.03},0.15]$ &  $[0, \textbf{3}, 5]$ \\
AUS   &   $120^\circ$ E    & $20^\circ$  S   &  $150^\circ$  E   &  $5^\circ$  N      & $[-170,\textbf{10},210]$ & $[-0.03,\textbf{0.05},0.27]$& $[-4,\textbf{0}, 4]$\\
SAM  &    $70^\circ$ W     & $20^\circ$ S    & $40^\circ$ W      &  $5^\circ$ S       & $[-200,\textbf{-70},160]$  & $[0,\textbf{0.12},0.28]$&$[0, \textbf{2}, 8]$ \\
NAM  &    $115^\circ$ W   & $20^\circ$ N    & $102^\circ$ W    & $35^\circ$ N      &  $[-20,\textbf{-60},-180]$    & $[0.07,\textbf{0.13},0.4]$ &$[-7,\textbf{1}, 10]$\\
 \bottomrule
\hline
\end{tabular}
\end{center}
\end{table}

\clearpage


\begin{figure*}
\centering
 \subfigure{ \includegraphics[trim = 0mm  0mm 0mm  50mm,clip,angle=0,width=1.2\textwidth]{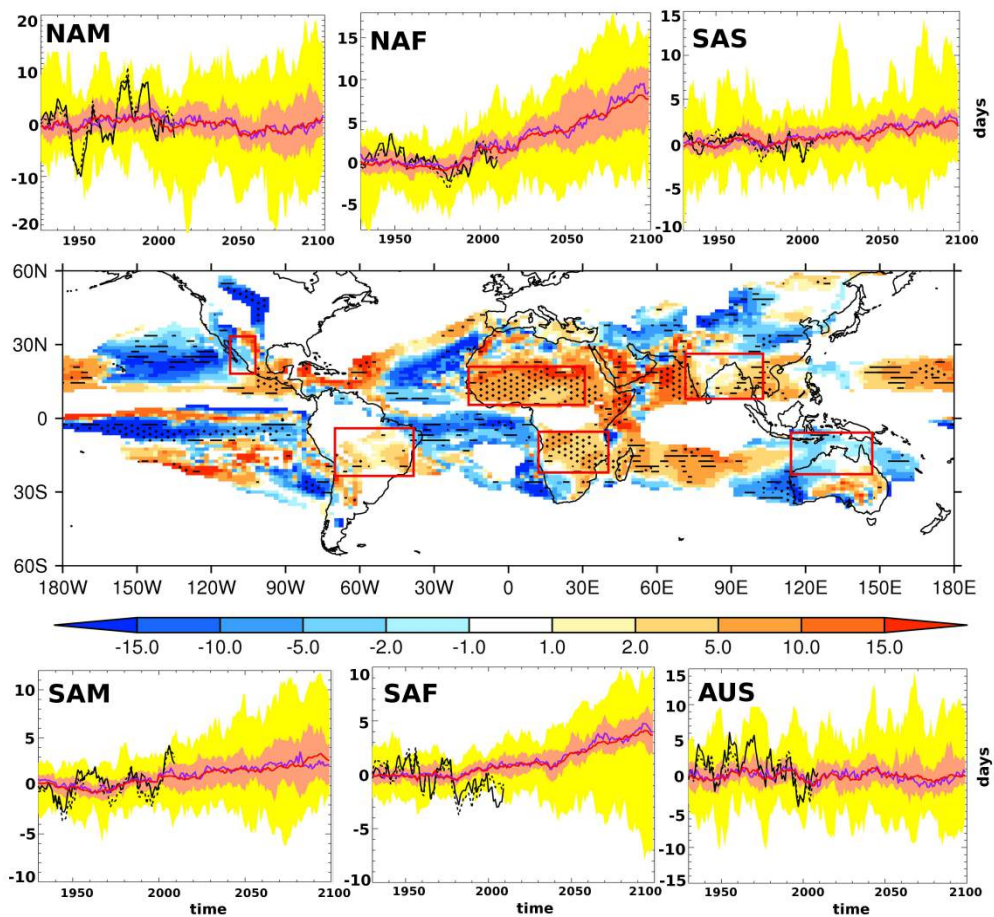} }
 \caption{ Delay (in days) of the rainfall  centroid  (2080-2100 and 1980-2000,  RCP8.5 scenario). Results are shown   just for areas with relative entropy   larger than $0.3$ in order to exclude regions without a pronounced dry season. Positive numbers  indicate a delay. Areas where  at least $90 \%$ of the models agree on the sign of the change are marked with stippling  while  areas where less than $66 \%$ of the models agree in sign are marked with horizontal hatching.  Time series (10-year running means) show the area-averaged    centroid delays  for the NAM, SAM, NAF, SAF, SAS, AUS monsoonal regions. The median (purple line), the total inter-model spread (yellow shading), the 25th and 75th percentiles (salmon shading) and the MEM (red line) of centroid delays  among the 24 models are shown. Results from precipitation gridded datasets  GPCC, black solid line,  and  CRU, black dashed line,  are also shown for comparison over the period 1930-2005.} 
\label{delay}      
\end{figure*}

\begin{figure*}
\centering
\subfigure{\includegraphics[trim = 20mm  0mm 0mm  0mm,clip,angle=-0,width=1.3\textwidth]{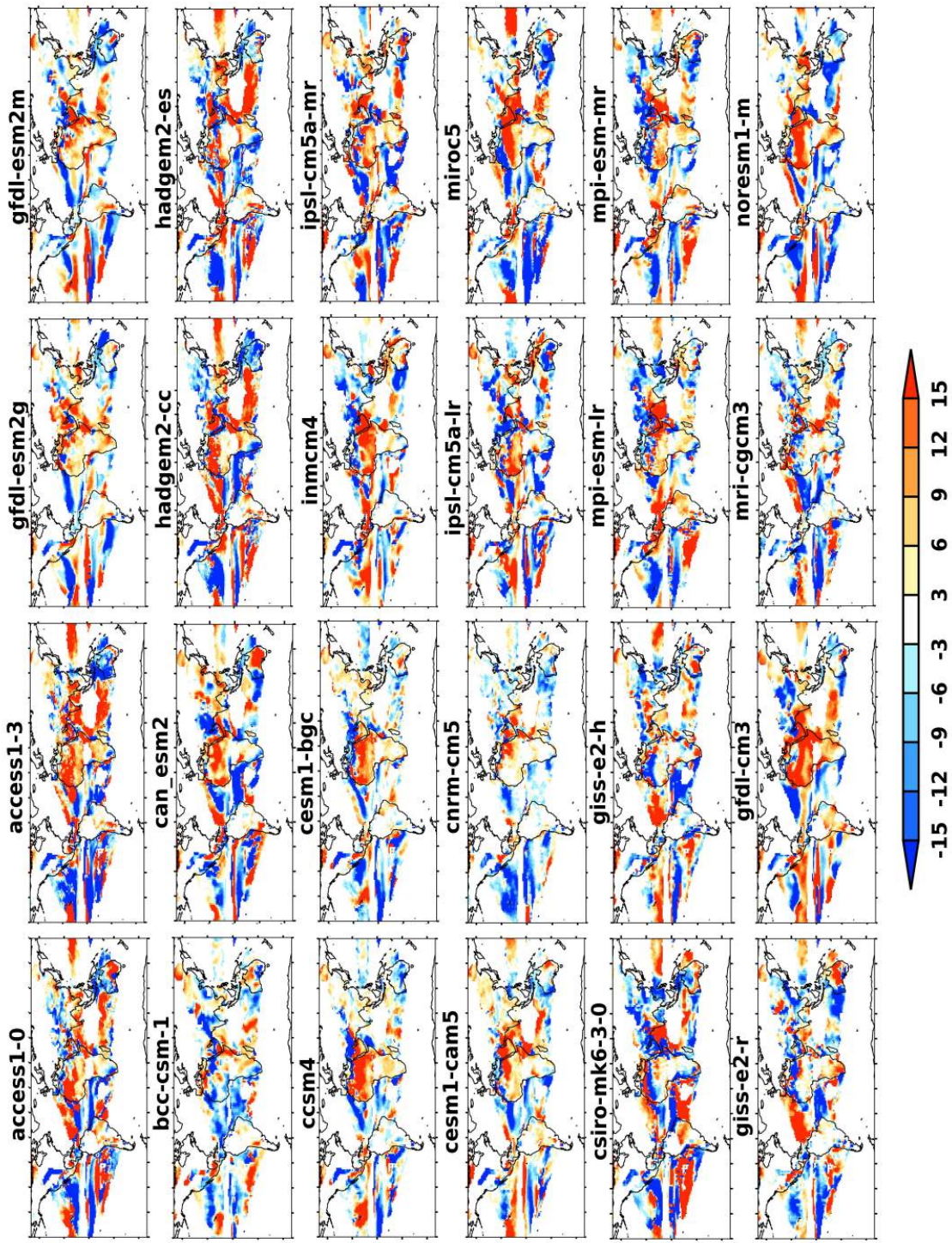} }
\caption{Delay (days) of the precipitation centroid   for the last 20 years of the twenty-first century (RCP8.5) minus last 20 years of twentieth century.}
\label{pannel2}       
\end{figure*}

\clearpage


%

%
%

\end{document}